\documentclass[
 reprint, 
superscriptaddress,
 amsmath,amssymb,
 physrev,
]{revtex4-2}

\usepackage{graphicx}
\usepackage{dcolumn}
\usepackage{bm}

\begin{document}


\title{\textbf{Adaptive Inference and Convergence of Free Energy Landscapes Using Non-parametric Bayesian Enhanced Sampling }}

\author{Daisy Kamp}
\affiliation{%
 Department of Materials Science and Engineering,
 University of California, Irvine, CA, 92697
}%

\author{Sinai Lee}%
\affiliation{%
Department of Physiology and Biophysics,
 University of California, Irvine, CA, 92697
}%
\author{Ronald Phung}
\affiliation{%
 Department of Materials Science and Engineering,
 University of California, Irvine, CA, 92697
}%
\author{Xavier Garcia}
\altaffiliation{%
Present address: Department of Computer Science,
California Polytech State University, San Luis Obispo, CA 93407
}%
\affiliation{%
 Department of Materials Science and Engineering,
 University of California, Irvine, CA, 92697
}%
\author{Joni Spencer}
\affiliation{%
 Department of Materials Science and Engineering,
 University of California, Irvine, CA, 92697
}%
\author{Alvin Yu}%
\email{Corresponding authors: alviny6@uci.edu, elizabeth.lee@uci.edu}
\affiliation{%
Department of Physiology and Biophysics,
 University of California, Irvine, CA, 92697
}%
\author{Elizabeth M.Y. Lee}%
\email{Corresponding authors: alviny6@uci.edu, elizabeth.lee@uci.edu}
\affiliation{%
 Department of Materials Science and Engineering,
 University of California, Irvine, CA, 92697
}%
\affiliation{%
 Department of Chemical and Biomolecular Engineering,
 University of California, Irvine, CA, 92697
}%
\date{\today}

\begin{abstract}
Enhanced sampling techniques are central to the study of statistically rare events in the computer modeling and simulation of molecular phenomena. In this work, we report the development and integration of a Gaussian process model, adaptive, uncertainty-driven sampling scheme for enhanced sampling. The framework trains a Gaussian process model on an iteratively improving estimate of the free energy landscape. Regions of high uncertainty within the reaction phase space are increasingly sampled, and the uncertainty is computed on-the-fly during free energy reconstruction, serving as a convergence metric. This approach provides a generalizable strategy that can be extended to complex molecular processes.

\end{abstract}

\maketitle

Free energy landscapes are fundamental descriptors of molecular systems that map the thermodynamically stable states and the underlying energetic barriers that govern the transitions between them. Consequently, they are essential for characterizing complex chemical events like bond-breaking, conformational switching, and phase transitions  \cite{lee_stability_2021,lee_neural_2021, dinner_understanding_2000, yu_molecular_2016, yu_energetics_2017, yu_atomic-scale_2020,yu_neurotransmitter_2018, santos_molecular_2025, cannistra_atomistic_2026, zhang_phase_2021}. To track the progress of these transitions, collective variables (CVs) or order parameters are typically defined, either from physical intuition or by identifying the parameters that best distinguish the relevant end states of the process~\cite{rogal_neural-network-based_2019,sidky_machine_2020, bonati_unified_2023,fu_collective_2024}. The free energy is then measured or computed as a function of these CVs.

These, free-energy landscapes can be reconstructed from force measurements or configurational probabilities along the selected reaction coordinates. Single-molecule pulling and fluorescence experiments have revealed the free-energy landscapes of conformational changes in proteins ~\cite{tapia-rojo_enhanced_2023,franceschini_force_2026,feng_cooperative_2026,modak_single-molecule_2024}, while colloidal free energies have been determined by total internal reflection microscopy and atomic force microscopy~\cite{cui_comprehensive_2022, vialetto_effect_2024}. In the computer simulation of molecular phenomena, enhanced sampling techniques--including umbrella sampling, metadynamics, and adaptive biasing force (ABF)  methods~\cite{kumar_weighted_1992, roux_calculation_1995,laio_escaping_2002,darve_adaptive_2008} that bias trajectories to accelerate sampling of rare transitions can be used to study statistically rare events. However, conventional methods rely on \textit{a priori} knowledge of the underlying energetics, which must be encoded via empirically determined parameters. For instance, the force constants of the biasing potentials in umbrella sampling determine the width and overlap of the sampled windows, whereas the height and tempering factor of the deposited Gaussian biases in well-tempered metadynamics~\cite{barducci_well-tempered_2008} determine the scheme's convergence and energy resolution. Gaussian process (GP) regression models are an ideal candidate for treating unknown functions as stochastic fields, defining a probability distribution over functions rather than a single point estimate~\cite{rasmussen_gaussian_2008}. As a non-parametric Bayesian model, the posterior can be updated on-the-fly using simulation statistics and iteratively refined with additional sampling~\cite{deringer_gaussian_2021, 
wee_quantification_2019-1, vandermause_--fly_2020, harris_active_2025}. Hence, there is no need to determine empirical fitting parameters, yielding a continuous posterior function that interpolates between sparse data points while providing rigorously quantified uncertainty~\cite{rasmussen_gaussian_2008}. These properties make GP models particularly well-suited for adaptive enhanced sampling.

Here, we present an adaptive, generalizable inference-based sampling framework that integrates GP models with enhanced sampling, leveraging the GP posterior distribution for simultaneous free-energy reconstruction and uncertainty-guided sampling. The method presented operates on-the-fly during simulation: mean forces are collected in real time, learned by a GP model, and simultaneously used to reconstruct a continuous free energy landscape and compute biasing forces that drive trajectories toward undersampled regions of phase space. Furthermore, the posterior uncertainty serves as a convergence criterion, enabling the framework to self-terminate. This combination of sparse force interpolation, adaptive uncertainty-driven sampling, and automated convergence provides a new framework that leverages the strengths of Bayesian inference for enhanced sampling.

In this Letter, we begin with a description of the general theory of the framework, followed by three applications. First, we demonstrate the scheme on a model 1D double-well potential. Second, we apply the GP-based sampling algorithm to identify the conformational changes in a benchmark biomolecular system, alanine dipeptide. Finally, we employ \textit{ab initio} molecular dynamics (AIMD) to map the free energy landscape of butane hydrogenolysis on a ruthenium catalyst. Collectively, these applications demonstrate that GP-based enhanced sampling robustly and efficiently uncover the free energy landscapes of complex molecular transitions.

\begin{figure}
\includegraphics[width=0.4\textwidth]{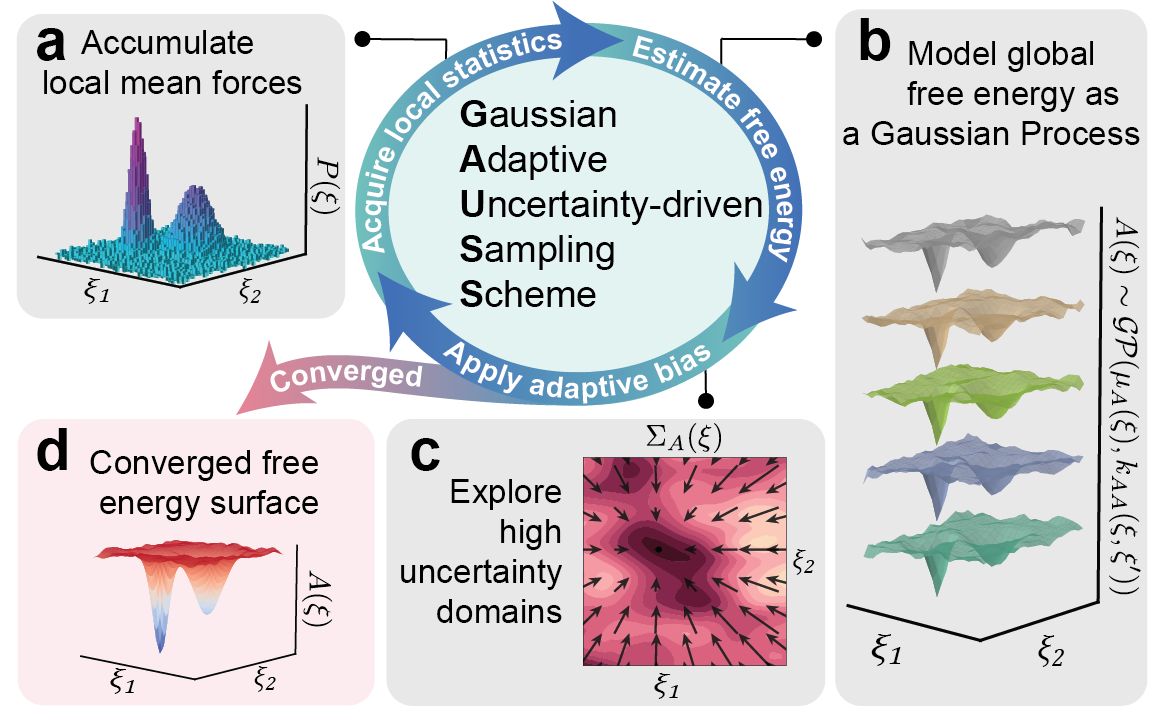}
\caption{\label{fig:1} Adaptive free energy sampling by GAUSS using Bayesian inference: (a) Local thermodynamic forces are sampled along predefined collective variables ($\boldsymbol{\xi}$) and (b) used as inputs to train a Gaussian process ($\mathcal{GP}$) to simultaneously infer the free energy $A(\boldsymbol{\xi})$, its associated bias forces, and its posterior variance $\Sigma^2_A(\boldsymbol{\xi})$. (c) High-uncertainty regions trigger an additional exploratory bias, and (d) accelerate sampling until convergence is reached.}
\end{figure}

Understanding complex molecular systems often requires mapping high-dimensional atomic coordinates, $\mathbf{r} \in \mathbb{R}^{3M}$, onto a lower-dimensional space of $d$ reaction coordinates or collective variables (CVs), $\boldsymbol{\xi}(\mathbf{r}) \in \mathbb{R}^d$, that characterize the reaction coordinates of interest. Thegradient of the free-energy surface $A(\boldsymbol{\xi})$ is related to the local mean force, $\langle \mathbf{f} \rangle_{\boldsymbol{\xi}} = -\nabla_{\boldsymbol{\xi}} A(\boldsymbol{\xi})$, obtained by ensemble-averaging instantaneous thermodynamic forces from the rate of change in atomic momenta along a trajectory~\cite{darve_adaptive_2008,guo_adaptive_2018}.

In conventional MD simulations, high free-energy barriers can trap the system in local minima, preventing exploration of the full $\boldsymbol{\xi}$ space. Overcoming these barriers requires applying an opposing biasing force, $\mathbf{f}_{\text{bias}}(\boldsymbol{\xi})$, which cancels the underlying physical forces. Because $A(\boldsymbol{\xi})$ is initially unknown, the bias must be estimated iteratively on-the-fly during the simulation. However, existing history-dependent biasing approaches converge slowly over large barriers when sparse sampling yields noisy, and therefore inaccurate, force estimates~\cite{invernizzi_exploration_2022}.

The framework, which we call the Gaussian Adaptive Uncertainty-driven Sampling Scheme (GAUSS), models the unknown free energy surface non-parametrically as a GP:
\begin{equation}
    A(\boldsymbol{\xi}) \sim \mathcal{GP}\big(\mu_0(\boldsymbol{\xi}),\, k_{AA}(\boldsymbol{\xi}, \boldsymbol{\xi}')\big).\label{eq:GP}
\end{equation},
which defines a prior probability distribution $p(A)$ over all possible free energy functions before any simulation data is observed. Here, $\mu_0(\boldsymbol{\xi})$ is the prior mean, taken as zero in the absence of \textit{a priori} knowledge; and $k_{AA}(\boldsymbol{\xi},\boldsymbol{\xi}')\big)$ is the covariance kernel that defines how input data $\boldsymbol{\xi}$ and $\boldsymbol{\xi}'$ relate to one another and determine the smoothness and uncertainty of the inferred free energy landscape. The hyperparameters $\boldsymbol{\Theta}$ of the kernel depend on the chosen functional form (e.g., radial basis functions), but they typically characterize the energy fluctuation scale ($\alpha$) and the spatial correlation length ($\ell$).

As the simulation proceeds, the GP model is updated, or conditioned, using the sampled CV configurations, $\boldsymbol{\Xi} = \{\boldsymbol{\xi}_1, \dots, \boldsymbol{\xi}_N\}$, and the corresponding aggregated mean forces, $\mathbf{F} = [\langle \mathbf{f}(\boldsymbol{\xi}_1) \rangle^T, \dots, \langle \mathbf{f}(\boldsymbol{\xi}_N) \rangle^T]^T$. Applying Bayes' theorem updates the prior to the posterior distribution $p(A \mid \mathbf{F}, \boldsymbol{\Theta})$, where the likelihood, $p(\mathbf{F} \mid A, \boldsymbol{\Theta})$, describes the probability of observing the sampled mean forces for a proposed free-energy surface $A$, while enforcing the physical relation $\langle \mathbf{f} \rangle_{\boldsymbol{\xi}} = -\nabla_{\boldsymbol{\xi}} A$. Because the derivative of a GP remains a GP~\cite{rasmussen_gaussian_2008}, the global free energy surface can be obtained analytically from local force measurements within the same probabilistic framework, without requiring numerical integration. Conditioning the GP on the accumulated simulation data yields the closed-form, nonparametric expressions for the inferred posterior mean, $\mu_A(\boldsymbol{\xi})$,
\begin{equation}
    \mu_{A}(\boldsymbol{\xi}) = \mu_{0}(\boldsymbol{\xi}) + \mathbf{k}_{A\mathbf{F}}\mathbf{K}_y^{-1}\big(\mathbf{F} - \mu_{\mathbf{F}}(\boldsymbol{\xi}) \big) 
    \label{eq:mu}
\end{equation}
and posterior variance, $\Sigma^2_A(\boldsymbol{\xi})$, 
\begin{equation}
    \Sigma^2_{A}(\boldsymbol{\xi}) = k_{AA}(\boldsymbol{\xi}, \boldsymbol{\xi}') - \mathbf{k}_{A\mathbf{F}}\mathbf{K}_y^{-1}\mathbf{k}_{A\mathbf{F}}^T \label{eq:sigma}
\end{equation}
which respectively provide the inferred free energy landscape and its uncertainty quantification. Here, $\mathbf{K}_{y} = \mathbf{K}_{\mathbf{FF}} + \sigma_n^2 \mathbf{I}$ represents the regularized force covariance matrix, where $\mathbf{K}_{\mathbf{FF}}$ is the covariance of the sampled mean forces and $\sigma_n^2$ is the observation-noise variance~\cite{solak_derivative_2002}. The kernel hyperparameters are iteratively optimized by maximizing the log-marginal likelihood~\cite{rasmussen_gaussian_2008}. Additional derivations are provided in the Supplemental Materials.

GAUSS directly maps the inferred posterior mean to the adaptive bias, $\mathbf{f}_{\text{bias}}(\boldsymbol{\xi}) = -\nabla_{\boldsymbol{\xi}} \mu_A(\boldsymbol{\xi})$, while the posterior variance $\Sigma^2_A(\boldsymbol{\xi})$ quantifies the remaining statistical uncertainty, enabling concurrent free-energy estimation and uncertainty quantification within a single simulation run. This on-the-fly uncertainty metric identifies undersampled regions in CV space, biasing the trajectory via an additional exploratory harmonic force $\mathbf{f}_{\text{ex}}(\boldsymbol{\xi})$. The total history-dependent biasing force is thus:
\begin{equation}
\mathbf{f}_{\text{bias}}(\boldsymbol{\xi}) = \nabla_{\boldsymbol{\xi}} \mu_A(\boldsymbol{\xi}) + \mathbf{f}_{\text{ex}}(\boldsymbol{\xi}).
\label{eq:total_bias}
\end{equation}
Algorithmically, GAUSS operates iteratively (Fig.~\ref{fig:1}), where the simultaneous inference of the free energy $\mu_A(\boldsymbol{\xi})$ and its posterior variance $\Sigma^2_A(\boldsymbol{\xi})$ determines the sampled trajectory. This cycle proceeds as follows: (i) the instantaneous thermodynamic force $\mathbf{f}$ is mapped to its collective-variable configuration and appended to the data matrix  $\{\mathbf{F}, \boldsymbol{\Xi}\}$ (Fig.~\ref{fig:1}a)); (ii) the hyperparameters $\boldsymbol{\Theta}$ are periodically optimized to update the global free energy and uncertainty models (Fig.~\ref{fig:1}b); and (iii) the continuous, history-dependent biasing force $\mathbf{f}_{\text{bias}}(\boldsymbol{\xi})$ from Eq.~\ref{eq:total_bias} is evaluated to flatten the underlying energy barriers (Figs.~\ref{fig:1}c,d). Importantly, GAUSS systematically drives additional sampling in regions of high $\Sigma_A$ until the average uncertainty falls below a threshold, ensuring an accurate free energy landscape.

\begin{figure}
\includegraphics[width=0.4\textwidth]{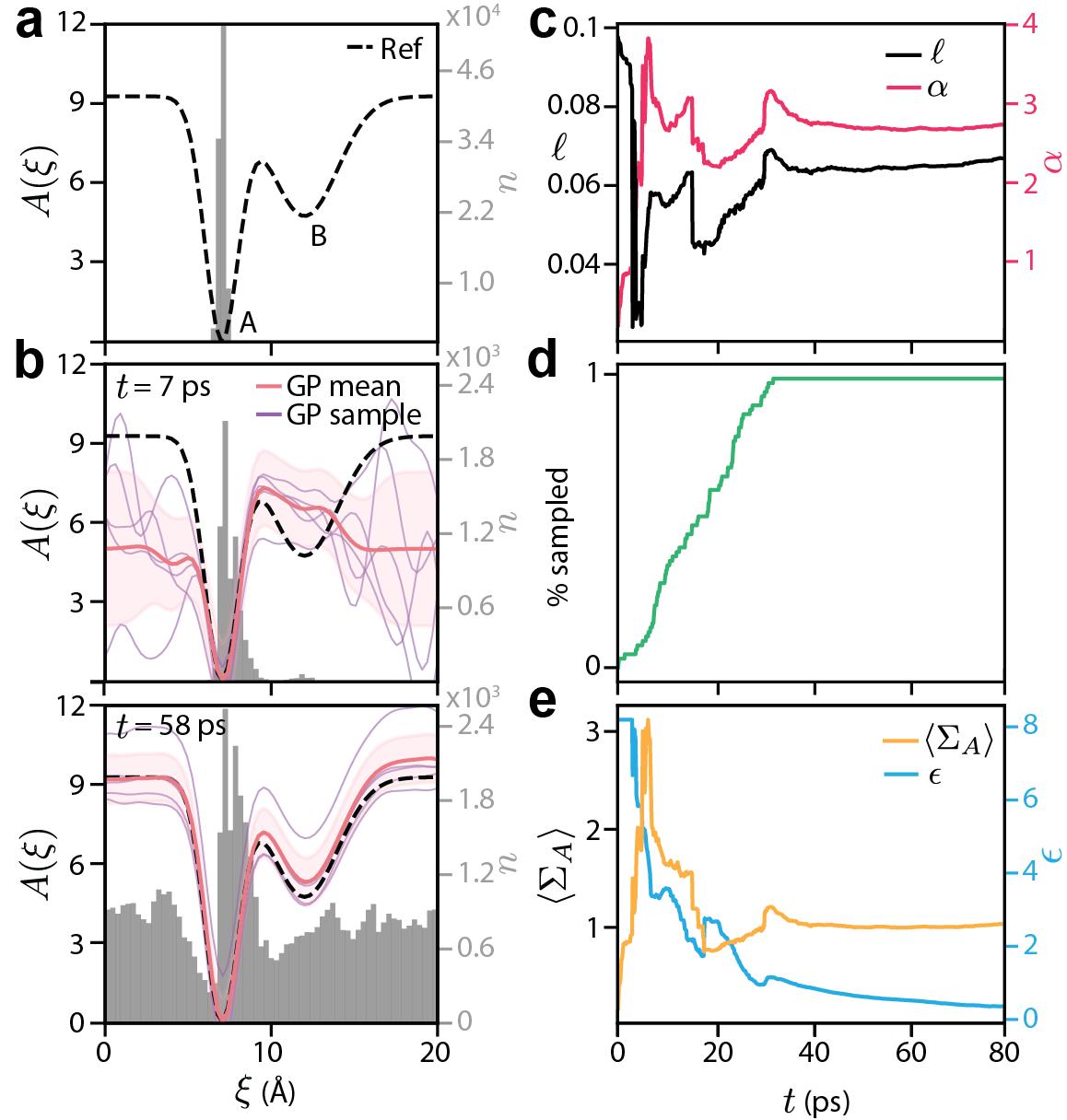}
\caption{\label{fig:2} Adaptive free-energy convergence on a 1D double-well potential. Energies and errors are normalized by $k_\text{B} T$ at $T=550\text{ K}$. (a) Unbiased 1 ns trajectory histogram ($n$, right axis) showing a system trapped in a single basin of the reference potential (dashed line, left axis). (b) Evolution of the inferred free-energy posterior mean $\mu_A(\boldsymbol{\xi})$ (solid line) and uncertainty $\Sigma_A(\boldsymbol{\xi})$ (shaded region), overlaid with samples from the posterior distribution (purple lines). Convergence is reached near $t = 58\text{ ps}$. (c) Dynamic evolution of optimized hyperparameters: length scale $\ell$ (normalized by CV range) and amplitude $\alpha$ (normalized by $k_\text{B} T$). (d) Fraction of configurational space $p(n)$ sampled at least $n_{\text{min}}=100$ times. (e) Mean posterior uncertainty $\langle \Sigma_A \rangle$ (left) and root-mean-squared error $\epsilon$ relative to the analytical reference (right) converging to $\sim 1 k_\text{B} T$ and below the thermal energy scale, respectively.}
\end{figure}

To demonstrate the adaptive sampling and convergence capabilities of GAUSS, we first consider a classic rare-event problem: a particle undergoing Langevin dynamics within a 1D double-well potential (\ref{fig:2}a). The model potential represents a chemical reaction between two states separated by an activation barrier of approximately 7 $k_\text{B}T$ or 0.38 eV at $T$ = 500 K (see Supplemental Materials for details of the simulations). 

GAUSS reconstructs the free-energy landscape by inferring the spatial correlations in free energies between configurations within the reactive phase space by training a GP model with measurements of local thermodynamic forces. The resulting non-parametric model extrapolates into unsampled regions in the configurational space, generating a thermodynamic force that counteracts the underlying forces at the boundary of the sampled region. We find that this biasing force drives the particle across the activation barrier within 7 ps (Fig.~\ref{fig:2}b, top). As additional force data are collected, the posterior mean rapidly converges toward the model potential while the posterior uncertainty contracts, illustrated by the narrowing distribution of free energy functions sampled from the GP posterior distribution. By 16 ps, the overall shape of the double-well potential is recovered, and by 58 ps, the particle undergoes a diffusive random walk after sampling the entire configurational space (Fig.~\ref{fig:2}b, bottom), indicating that the computed free energy landscape is converged (see also Supplemental Movie S1).

The evolution of the optimized hyperameters, $\boldsymbol{\Theta} = \lbrace \alpha,\ell \rbrace$, provides a quantitative measure of structural changes in the inferred free energy landscape over time (Fig.~\ref{fig:2}c). Peaks during the first tens of picoseconds coincide with visits to previously unsampled regions. The final, smaller peak occurs at 35 ps, when a nearly complete convergence of the configurational space is achieved, defined here by each CV state being sampled at least a hundred times (Fig. 2d). Beyond this point, the hyperparameters remain relatively unchanged despite continued sampling, indicating that the GP model no longer acquires new samples that would change the covariance structure. This behavior suggests an intrinsic self-termination criterion once the free energy landscape has been fully recovered.

The posterior uncertainty provides a complementary convergence metric. Before crossing the barrier, the GP posterior has a large uncertainty, $\langle\Sigma_A\rangle \approx 3\,k_B T$. Following complete exploration of the CV space, the uncertainty contracts to $\langle\Sigma_A\rangle \approx 1\,k_B T$ at 58 ps (Figs.~\ref{fig:2}d,e), coincident with the root-mean-squared error, $\epsilon$, of only 0.02 eV ($\approx 0.4 k_\text{B}T$) relative to the model potential (Fig.~\ref{fig:2}e). Together with the complete configurational-space coverage and stabilization of the GP hyperparameters, the posterior uncertainty establishes a general, self-consistent convergence criterion for GAUSS.

To evaluate GAUSS on a molecular system, we apply the method to alanine dipeptide in water, a notable benchmark system for enhanced sampling methods whose conformational dynamics are captured by a rugged two-dimensional free energy landscape spanned by the backbone dihedral angle $\phi$ and $\psi$~\cite{yu_computing_2016} (see Supplementary Materials for simulation details).

\begin{figure*}
\includegraphics[width=0.8\textwidth]{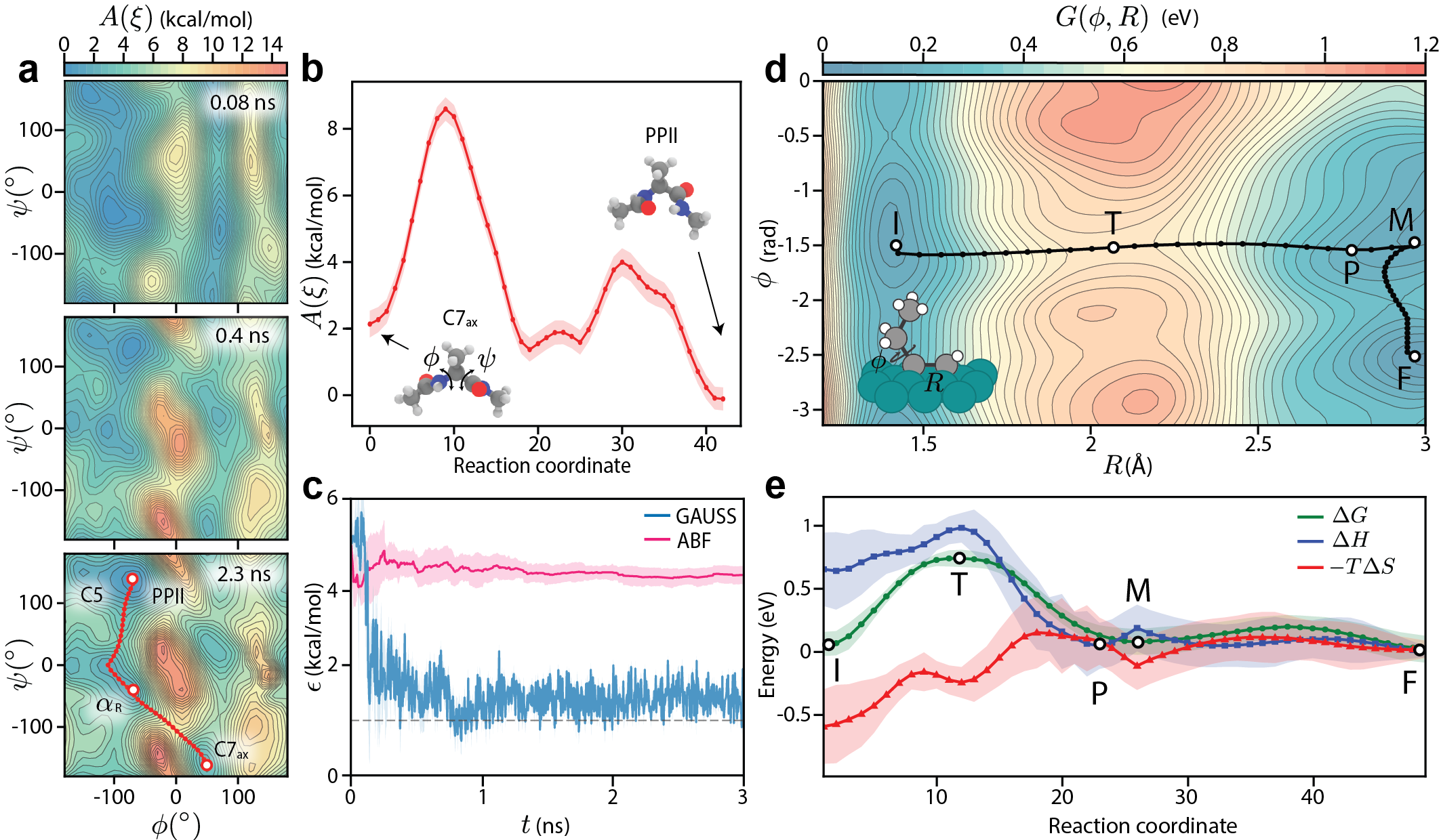}
\caption{\label{fig:3} 
Free energy reconstruction of molecular rare events using GAUSS. (a) Evolution of the free energy landscape of alanine dipeptide over time. The minimum free-energy pathway (MFEP) is shown in red in the final panel. (b) Free energy profile along the MFEP for the $C7_\text{ax}$ to PPII transition. (c) Root-mean-squared error $\epsilon$ relative to the reference free-energy profile, showing that GAUSS converges within 2$k_{\text{B}}T$ (dashed line) and outperforms the adaptive biasing method. (d) Free energy landscape of catalytic C-C bond cleavage of  butane on Ru(0001), showing the initial (I), transition (T), local potential-energy minimum (P), metastable (M), and final dissociated (F) states. The MFEP (black) reveals a two-step bond cleavage mechanism. (e) 1D free energy profile along the MEFP with free energies decomposed into enthalpic $(\Delta H)$ and entropic $(-T\Delta S)$ contributions.}
\end{figure*}

Among its metastable conformations, the $C7_\text{ax}$ $\rightarrow$ PPII transition is the slowest because it requires coordinated changes in both dihedral angles~\cite{best_optimization_2012,pane_development_2021}. In Figure~\ref{fig:3}a, we show the evolution of the inferred free-energy landscape. The overall topology is recovered early in the simulation at around 0.4 ns and progressively refined, converging after 2.3 ns once the configurational-space coverage, hyperparameter stabilization, and posterior uncertainty simultaneously converge (see Fig. S4). The converged free-energy surface accurately reproduces the $C7_\text{ax}$ and PPII energy minima at $(50^\circ, -160^\circ)$ and $(-70^\circ, 140^\circ)$, respectively, consistent with prior studies.~\cite{best_optimization_2012,pane_development_2021,cruz_water-mediated_2011, garcia-prieto_study_2011} (see Supplemental Movie S2 for the alanine dipeptide simulations trajectory).

The posterior variance identifies undersampled regions within the backbone-dihedral configurational landscape and adaptively steers sampling toward configurations of highest uncertainty (see Fig. S1). After 2.3 ns of simulation, the mean uncertainty decreases to 0.58 kcal/mol, comparable to the thermal energy of $1 k_\text{B}T$ (0.59 kcal/mol at 300 K), indicating convergence of the free-energy reconstruction.

The converged free-energy surface reveals a minimum free-energy pathway connecting the $C7_\text{ax}$ and PPII conformations through two intermediate states, with a free energy barrier of 6.12 kcal/mol (Fig.~\ref{fig:3}b). To examine how GAUSS accelerates free energy calculations, we also performed simulations using the adaptive biasing force (ABF) method. For both GAUSS and ABF, we computed the root-mean-square error along the free energy profile against a reference from umbrella sampling simulations (see Supplementary Materials for details). GAUSS achieves a root-mean-square error of only 1.7 kcal/mol, compared with approximately 4 kcal/mol for ABF (Fig.~\ref{fig:3}c). Notably, the GAUSS error falls to nearly $2 k_\text{B}T$, or 1.2 kcal/mol, indicating accurate reconstruction of the free-energy landscape within thermal noise. These results demonstrate that Bayesian inference of the free-energy surface, coupled with uncertainty-guided exploration, accelerates convergence while preserving quantitative accuracy.

We next apply GAUSS to catalytic carbon–carbon (C-C) bond cleavage during alkane hydrogenolysis, a key elementary step in the chemical upcycling of polyolefin plastics~\cite{rorrer_conversion_2021, rorrer_hydrogenolysis_2021}. Conventional free-energy calculations based on density functional theory (DFT) rely on harmonic approximations of predetermined minima and transition states~\cite{almithn_comparing_2019, hibbitts_effects_2016}, limiting the ability to capture complex reaction energy landscapes with configurational entropy and anharmonic effects. Here, we directly reconstruct the two-dimensional free-energy landscape of butane bond cleavage on Ru(0001) from DFT-based \textit{ab initio} molecular dynamics (AIMD) using the C–C bond distance, $R$, and the backbone dihedral angle, $\phi$, as collective variables (see Supplementary Materials for simulation details).

The free-energy landscape converges within 11.7 ps using ten simulation replicas (Fig.~\ref{fig:3}d; see also Supplemental Movie S3). The minimum free-energy pathway reveals a two-step mechanism in which C–C bond stretching first overcomes a free-energy barrier of 0.45 eV to form a metastable intermediate (state M), followed by backbone rotation into the final dissociated state (state F), producing a three-carbon fragment and an adsorbed CH species (see Fig. S4). The lower barrier relative to previous estimates for ethane highlights the influence of molecular size and conformational flexibility on catalytic bond cleavage~\cite{almithn_comparing_2019} (see Supplemental Movie 3 for butane backbone rotation and cleavage). Moreover, the two-step mechanism suggests that, following bond cleavage, the adsorbed fragments first relax to favorable conformations before undergoing dehydrogenation and desorption during hydrogenolysis.

Decomposition of the free energy into enthalpic and entropic contributions (Fig.~\ref{fig:3}e) reveals that bond stretching is accompanied by a substantial loss of configurational entropy, while electron transfer from the Ru surface into C–C antibonding orbitals weakens the bond and promotes dissociation (see Fig. S4). Interestingly, a shallow local minimum in the potential-energy surface (state P) appears adjacent to the metastable state in the free-energy surface. Although states P and M have nearly identical atomic structures (see Fig. S4), they occupy distinct regions of the free-energy landscape, indicating that thermal fluctuations and configurational sampling produce appreciable anharmonic contributions beyond the harmonic approximation. 

In summary, we have developed GAUSS, an inference-based enhanced sampling framework, that integrates Bayesian inference using a Gaussian Process model with adaptive sampling. By using local mean forces as the training data and applying additional exploratory biases guided by the posterior uncertainty, the framework drives the simulation toward unexplored regions of phase space. The posterior uncertainty is also used to quantitatively determine the convergence of the free energy landscape, enabling reliable free-energy calculations even for uncharacterized energy landscapes where reaction timescales are unknown. Across all benchmark systems studied here, the optimized Gaussian Process length scale consistently converged to approximately 10\% of the collective-variable range, suggesting a characteristic correlation length that may provide a practical heuristic for adaptive free-energy sampling.

The simultaneous inference of free-energy landscapes and their associated uncertainty quantification within a single simulation run makes GAUSS well-suited for large-scale molecular dynamics and \textit{ab initio} simulations, where obtaining statistically converged free-energy estimates remains computationally demanding. More broadly, this work demonstrates that statistical uncertainty in free energy can serve as both a quantitative convergence metric and an adaptive driving force for exploring complex free-energy landscapes, offering a general framework for studying rare molecular events, ranging from conformational transitions to chemical reactions. 

\textit{Acknowledgments}---This work was supported by the U.S. National Science Foundation under Grant No. CBET-2443952 (E.M.Y.L.) and the National Institute of Allergy and Infectious Diseases of the National Institutes of Health under Awards R00-AI167034 and U19-AI181968 (A.Y.). Computational resources were provided by the Texas A\&M University ACES cluster; Stampede3 at the Texas Advanced Computing Center (TACC) through the Advanced Cyberinfrastructure Coordination Ecosystem: Services \& Support (ACCESS) program under Allocations MAT250078 (E.M.Y.L) and BIO230061 (A.Y.), supported by NSF Grants Nos. 2138259, 2138286, 2138307, 2137603, and 2138296; the High Performance Computing Cluster (HPC3) at the Research Cyberinfrastructure Center (RCIC), University of California, Irvine; and the Carbon cluster at the Center for Nanoscale Materials (CNM Proposal No. 83620, E.M.Y.L.), a U.S. Department of Energy Office of Science User Facility, supported by the U.S. Department of Energy, Office of Basic Energy Sciences, under Contract No. DE-AC02-06CH11357. Anton 3 computer time was provided by the Pittsburgh Supercomputing Center (PSC) through NIH Grant R24-GM15042 under Allocation MCB240004P (A.Y.). The Anton 3 system at PSC was made available by D. E. Shaw Research.

\bibliographystyle{apsrev4-2}

\bibliography{GAUSS_bib_v4}

\end{document}